\documentclass[english]{svjour3}
\usepackage[utf8]{luainputenc}
\usepackage{url}
\usepackage{amsbsy}
\usepackage{amstext}
\usepackage{amssymb}
\usepackage{graphicx}

\makeatletter

\providecommand{\tabularnewline}{\\}

\RequirePackage{fix-cm}

\smartqed  

\makeatother

\usepackage{babel}
\begin{document}

\title{Dynamics of quantized vortices moving towards reconnection}

\subtitle{%
\thanks{This work was supported by RFBR grants No. 13-08-00673 and 15-02-05366.%
}}

\author{V.A. Andryushchenko \and  L.P. Kondaurova \and  S.K. Nemirovskii}

\institute{Institute of Thermophysics, Lavrentyev ave, 1, 630090, Novosibirsk,
Russia and Novosibirsk State University, Novosibirsk \email{vladimir.andryushchenko@gmail.com}}

\date{Received: date / Accepted: date}
\maketitle
\begin{abstract}
The main goal of this paper is to investigate numerically the dynamics of quantized
vortex loops, just before the reconnection at finite temperature, when mutual friction essentially changes evolution of lines.
Modeling is performed on the base of vortex filament method with using the full Biot-Savart
equation. It was discovered that initial position of vortices and the temperature strongly
affect the dependence on  time of the minimum distance $\delta
(t)$ between tips of two vortex loops.
In particular, in some cases the shrinking and collapse of vortex loops due to mutual friction occur earlier than reconnection, thereby cancelling the latter.
However, this relationship takes a universal square-root form $\delta\left(t\right)=\sqrt{\left(\kappa/2\pi\right)\left(t_{*}-t\right)}$ at distances smaller than the one,
satisfying the Schwarz criterion, when nonlocal contribution to the Biot-Savart equation becomes about
equal to local contribution. In the "universal" the nearest parts of vortices form a pyramid-like structure
with angles which are also don't depend both on the initial configuration of filaments and on the temperature.


\medskip{}

\end{abstract}

\noindent
\textbf{1 Introduction. Scientific background and motivation}

The reconnection of vortex  lines is of greatest influence on the dynamics of
quantum turbulence (QT) in superfluid helium \cite{key-1}. Thus, because
of reconnection processes, the vortex loops constituing QT permanently
 split and merge, forming the self-maintaining structure - the so called vortex tangle.
For this reason, the investigation of a reconnection of vortices is very important.
In particular, the dynamics of vortex loops just before
reconnection  is responsible for the rate of reconnections.
Also, an exact shape of lines (vortex line configuration) is crucial for the form
of spectrum of 3D motion, induced by these lines, etc.

Study of the time dependence of the minimum distance $\delta
(t)$ between tips of two vortex loops, as well as determination of the configuration
 of vortex lines  just before reconnection are the problems of special interest.

Of course, the problem of reconnection of quantized filaments was intensively
explored. Let us briefly describe the works that are the most relevant to our stated tasks. The
various  approaches are usually used for study of the reconnection problems. It is modeling, based on the Biot-Savart equation \cite{key-2,key-3,key-4}, and the numerical
solution of the Gross-Pitaevskii equation (GPE) \cite{key-4,key-5,key-6}. There are
also analytic studies (see, e.g. \cite{key-7,key-10}  and references therein). Besides, there are also
 experimental observations \cite{key-8}. In pioneering works \cite{key-2,key-10} it was found that at zero temperature the time dependence of
minimum distance between points of two vortex loops has a following
form \cite{key-2}:
\begin{equation}
\delta\left(t\right)=\sqrt{\left(\kappa/2\pi\right)\left(t_{*}-t\right)},
\end{equation}
where $\kappa=\text{h /\ensuremath{m_{He}}}$ is quantum of circulation,
$h$ is Planck's constant, $m_{He}$ is the mass of the helium atom,
$t_{*}$ is the reconnection time. It was also discovered that the geometrical configuration
of vortex lines at their route to the reconnection, forms an universal pyramidal
structure. That pyramidal
structure is independent on the initial position of vortex
loops \cite{key-2}.

However, in experimental work, performed
at  temperatures $\left(1.7\:K<T<2.05\:K\right)$
\cite{key-8} (and hence, the friction force was not zero) it was
obtained the following relation for quantity  $\delta
(t)$ :
\begin{equation}
\delta\left(t\right)=A\sqrt{\kappa\left(t_{*}-t\right)}\left(1+c|t_{*}-t|\right).
\end{equation}
Coefficients $A$ and $c$ were distributed around mean values $A\thickapprox1.25$,
$c\approx0.5s^{-1}$ with large dispersion. Remarkably that values of all $A$
were greater than $\sqrt{1/2\pi}$.

The described above discrepancy attracted attention of scientists, a series of works
on the dynamics of reconnecting lines was performed. The results of numerical simulations of the problem
obtained on the basis of the GPE \cite{key-4,key-5,key-6}, where many details of reconnection process
were refined, look promising. However, the dynamics of vortices in Bose-Einstein  gases is not fully equivalent to the dynamics of vortices in the He II.
The main sources of deviations are the large compressibility and the large core size.
Therefore, the difference (between GPE and vortex filament method (VFM)) would be significant,
up to that the square-root law would be strongly corrected for BEC.

The distance $\delta\left(t\right)$ was
studied in the approximation of the vortex filament method \cite{key-3}
at finite temperatures. It was found: $\delta\left(t\right)\sim\sqrt{\left(t_{*}-t\right)}$.
However, the information related to configuration of vortices before
reconnection wasn't disclosed there. Therefore the question of vortex dynamics just before
reconnection at finite temperature  in pure He II remains open.

In this paper we present the numerical results (which were based on VFM) of vortex loops dynamics
on their route to reconnection point at finite temperatures.\\

\noindent
\textbf{2 Basic equations and their implementation}

Numerical simulations were performed on the base of the VFM, solving the equation of motion for the vortex elements in a resting helium $\boldsymbol{V}_{ns}\left(s\right)=0$ :
\[
\boldsymbol{V}_{L}=\boldsymbol{V}_{B}-\alpha\boldsymbol{s}'\times\boldsymbol{V}_{B}
+\alpha'\boldsymbol{s}'\times\left[\boldsymbol{s}'\times\boldsymbol{V}_{B}\right].
\]
Here $\alpha,\:\alpha'$ are the mutual friction coefficients, $\boldsymbol{s}'$ is tangential vector.
Primes denote differentiation with respect to the instantaneous arc length. The $\boldsymbol{V}_{B}\left(s\right)$
is found by solving the Biot-Savart equation:
\begin{equation}
\boldsymbol{V}_{B}\left(\boldsymbol{s}\right)=\frac{\kappa}{4\pi}\int_{l}\frac{\left(\boldsymbol{s}_{1}-\boldsymbol{s}\right)\times d\boldsymbol{s}_{1}}{|\boldsymbol{s}_{1}-\boldsymbol{s}|^{3}},
\end{equation}
Here $\boldsymbol{s}_{1},\:\boldsymbol{s}$ are the
radius vectors of vortex points, the integration is taken over the
entire vortices configurations $l$. Initial configuration of system
was two rings of identical diameter lying in the same plane.The initial radius
of the vortex loops $R_{0}$ was varied from $10^{-5}m$ to $10^{-7}m$.
The initial distance between centers of rings was varied from $2.1\cdot R_{0}$
to $2.5\cdot R_{0}$. Starting from early works by Siggia \cite{key-9} and Schwarz \cite{key-10,key-11},
it was observed that the arbitrary oriented vortex lines,
approaching to each other, were changing their orientations to become anti-parallel. Thus, the  anti-parallel scenario is the most important and therefore, vectors of vortex rings circulation were chosen co
directional. The temperature of the system was varied  from $0\:K$ to $1.9\:K$, the corresponding friction coefficients were taken from \cite{key-12}.

The Runge-Kutta scheme of 4-th order was used to integrate the motion equation. Specific
details related to simulation algorithm and features of its numerical
implementation were described in \cite{key-13}. We varied space resolutions to test applicability of numerical algorithm
for investigated system. Discrepancies in the vortices velocities were less then
 one percent, for different spatial resolutions, but for identical initial conditions and temperatures.
Thus, the chosen numerical algorithm seems quite appropriate for investigation of vortex loops dynamics.

In the data presented in this work the second
friction coefficient $\alpha'$ was omitted. It was found that the influence of it was negligible, within  $1\div2$ percents.

In addition, the system was tested on the similarity
by studying of the rings of various initial radii. In principle the similarity would be
violated because of the logarithmic
factor. However, it was found that vortices evolve similarly (just before reconnection), despite
the presence of these factors.\\

\noindent
\textbf{3 Minimum distance between vortices}

Performed calculations illustrate that qualitatively  the evolution process is the
following:  due to non-local Biot-Savart  action the nearest parts go out of the
 initial plane with formation of cusps. After that the self-induced velocity, directed
 to opposite cusp, appears. Thus, the out-of-plane velocity is accompanied with the approaching factor.
As a result, the  close parts of both lines form a pyramid-like structure (see, Fig. 2(a)-(2b)). Qualitatively, this scenario is consistent with the results obtained by other researchers cited above. However, the mutual friction modifies an overall dynamics of approaching filaments.

\begin{figure}
\includegraphics[scale=0.35]{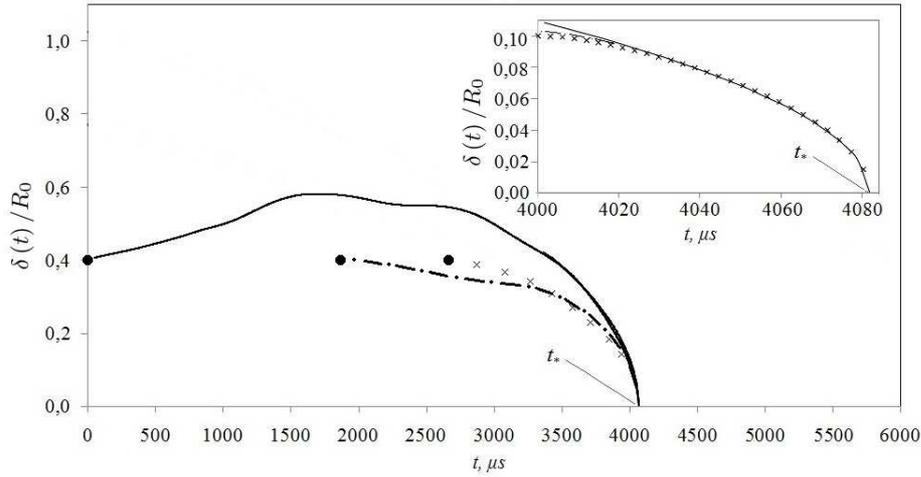}
\caption{ The time dependence of the quantity $\delta
(t)/R_{0}$. The solid line corresponds to $T=1.6\:K$, dash-dotted line corresponds to $T=1.3\:K$, crosses correspond to $T=0\:K$ (and obey equation (1)). All  initial distances are equal  $0.4\cdot R_{0}$.
(Inset) Increased region near the reconnection point.  }
\end{figure}
Let us consider such important question as a time dependence of the minimum distance between two vortex loops $\delta\left(t\right)$.  Hereinafter, we consider
the $R_{0}=10^{-5}m$ in most calculations.

In Fig. 1 the quantities $\delta(t)$ for three different temperatures $T=0\:K$,  $T=1.3\:K$, and $T=1.6\:K$ are depicted. In the all three cases, the initial normalized distances $\delta\left(0\right)/R_{0}$ are equal  $0.4$. The first remarkable, although not surprising, fact is that the reconnection times are different for all three cases. Therefore, in order to accomplish comparison, the curves for  temperatures $T=0\:K$ and  $T=1.3\:K$ are shifted along the time axis, so that the points of reconnections (in general, different) are coincided.

The shortest time for collision of filaments occurs at $T=0\:K$. The dependence $\delta\left(t\right)/R_{0}$, shown by crosses,  is well described by  square-root law (Eq.(1)). This result  is consistent with data of paper \cite{key-2}.  The reconnection time $t_{*}$ at zero temperature time depends
on the initial distance, and can be estimated as follows: $t_{*}=c[\delta\left(0\right)/R_{0}]^{2}$,
where $c\sim9.4\cdot10^{-3}s$.

Unlike the previous case, the curves  $\delta\left(t\right)$, obtained at finite temperatures,  behave in more sophisticated manner.
Conditionally, the whole evolution of colliding lines at finite temperatures consists of three parts. The third, last interval, just before reconnection, is fully universal. It is characterized by the square root dependence (on time) of shortest distance $\delta$ between tips on loops $\delta(t)\propto \sqrt{\kappa(t_{*}-t)}$ with the universal unchanged prefactor $\sqrt{1/2\pi}$. The second, next to last interval is semi-universal. The behavior of $\delta$  is close to the square root dependence, however, it is rather characterized by the corrected formula Eq. (2) with non-generic parameters $A$ and $c$,  and $A$ is greater than $\sqrt{1/2\pi}$. This results agrees well with the observation of the works  \cite{key-6,key-8}. Remarkably, the boundary between these two intervals (that was found numerically) is of the order of the Schwarz criterion for reconnection ansatz \cite{key-11,key-12}.  Finally, the first, initial part of evolution of vortices, before the semi-universal interval,  is non-universal at all. Its dynamics depends on many parameters (temperature, initial separation, radius of rings, etc.).\\

\noindent
\textbf{4 Shrinking and collapse of vortex loops}

With increasing of temperature (or mutual friction), the moving loops tend to shrink via friction.
As the result, under certain initial conditions, loops may collapse and die before reconnection.
This result, of course, can be predicted qualitatively. However, it would be useful to obtain more
or less rigorous quantitative criteria for this phenomena.
To estimate the characteristic times of the evolution of
loops (either reconnection or collapse time (lifetime)) we modeled system dynamics
at different initial temperatures and distances (see, Table 1).
\begin{center}
\begin{tabular}{|c|ccccc|}
\hline
T, K &  &  & $\delta\left(0\right)/R_{0}$ &  & \tabularnewline
\cline{2-6}
 & 0.1 & 0.2 & 0.3 & 0.4 & 0.5\tabularnewline
\hline
0.0 & 80.0 & 323.4 & 763.0 & 1410.7 & 2269.2\tabularnewline
1.3 & 80.9 & 366.1 & 1025.6 & 2184.3 & 3022.0\tabularnewline
\cline{6-6}
1.6 & 81.7 & 529.5 & 2490.3 & \multicolumn{1}{c|}{4064.8} & - \tabularnewline
\cline{3-5}
1.9 & \multicolumn{1}{c|}{84.0} & - & - & - & -\tabularnewline
\hline
\end{tabular}
\par\end{center}
\begin{center}
\textbf{Table 1:} Dependence of the collision time
(in microseconds) on temperature and initial distances between vortices.
\par\end{center}

The upper part of the table corresponds to a scenario where vortex
loops collide with each other, the lower, marked part corresponds
to scenario where the loops shrink and collapse  (due to mutual friction)
before reconnection.

For the temperatures equal to $1.3\, K$, $1.6\, K$, $1.9\, K$ and the initial radius
of the rings $R_{0}=10^{-5}m$ the lifetimes roughly equals to $15167\, \mu s$,
$5570\, \mu s$, $2600\, \mu s$, respectively. These values are of the order
of the lifetimes $\tau$ of  ideal vortex rings  $\tau=R_{0}^{2}/2\alpha\beta$, where $\beta=\left(\kappa/4\pi\right)\ln\left(R_{0}/a_{0}\right)$ and $a_{0}$ core radius (see, e.g.  \cite{key-14}).
It looks a bit surprisingly, since the vortex loops undergo to significant deformation due to  "tidal" non-local action during evolution.\\

\noindent
\textbf{5 Geometric configuration of loops before reconnection}

As mentioned, during evolution the closest sections of the loops form a pyramid structure as it
shown in Fig. 2(a).
\begin{center}
\includegraphics[scale=0.25]{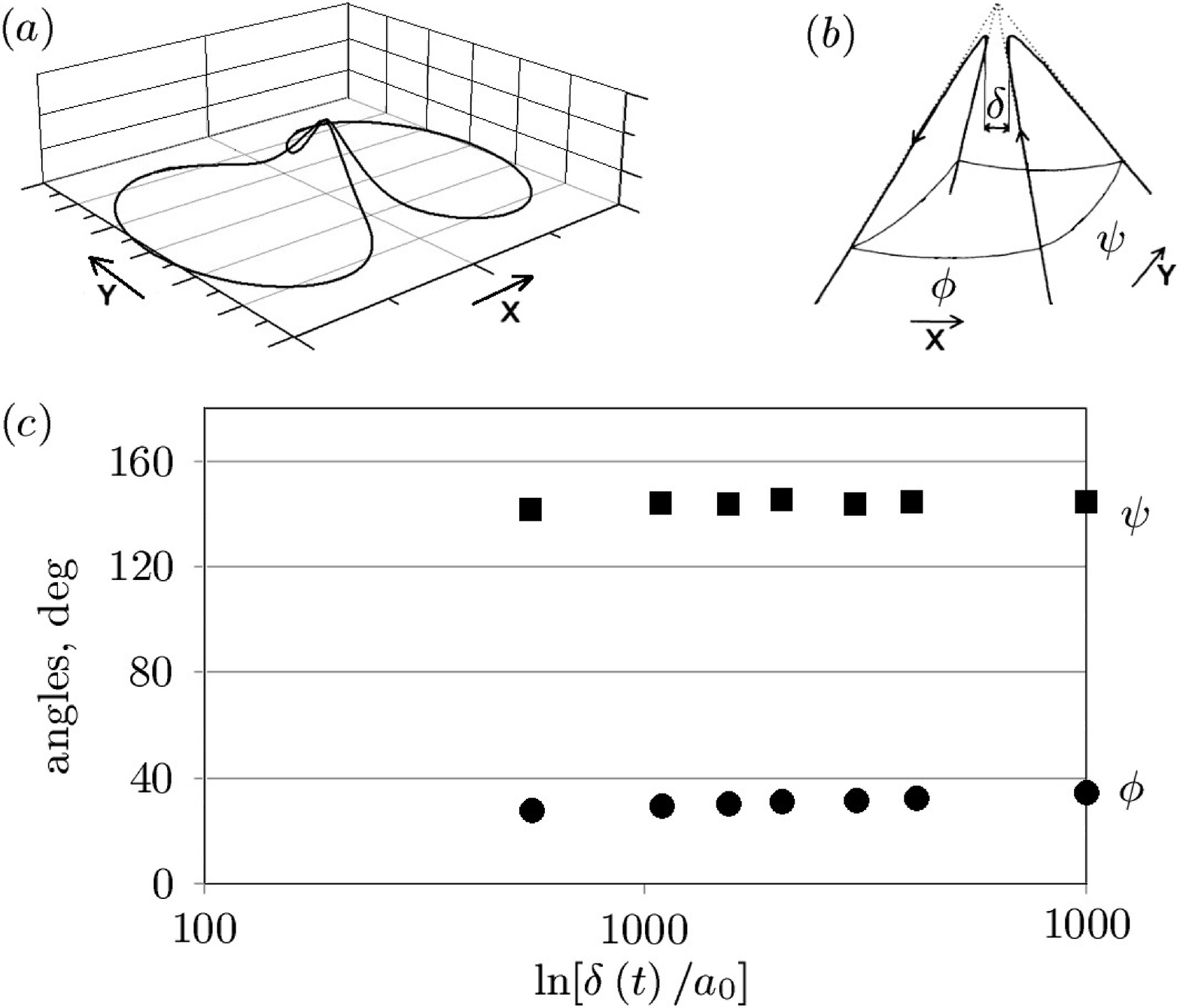}
\par\end{center}
\begin{center}
\textbf{Fig. 2} (a) Illustration of configuration of two vortex loops
before reconnection, (b) Definitions of angles, (c) Dependencies of angles on $\delta$ for $T=1.3\:K.$
\par\end{center}

Again, just as in the case of the minimum distance described in Sec. 4, there are also universal  and non-universal intervals. In the latter case, the influence of the mutual friction is significant. In the former case
the angles of the pyramid $\psi$ and $\phi$ practically
stay unchanged, which corresponds to the pattern, described in~\cite{key-2}.
We calculated the angles at very small distances, about several hundred of the core
size. In this situation the angles obtained at vertex
of pyramid equal to $\approx27$ and $\approx141$ degrees
accordingly, which were a little higher than values given in~\cite{key-2} as it
shown in Fig. 2(c). The angles at the top
of pyramid (immediately before reconnection) were practically independent
from initial positions of vortex loops and friction force, as it should be expected
in the universal regime.\\

\noindent
\textbf{6 Conclusion}

Thus, we revealed that the whole  evolution of colliding lines consists of three parts. These are the fully non-universal interval, depending on many parameters, and the final stage, which, in turn, is composed of two periods. The last interval, just before reconnection, is fully universal. It is characterized by the square root dependence  of shortest distance $\delta$ between tips on loops $\delta(t)\propto \sqrt{\kappa(t_{*}-t)}$ with the universal unchanged pre-factor $\sqrt{1/2\pi}$.  Another part of evolution of vortices, prior to universal  interval,  is semi-universal.  In particularly, its dynamics depends on the temperature, and the behavior of $\delta$  is characterized by the corrected formula Eq. (2). Therefore, any attempt to fit the function $\delta(t)$, obtained experimentally or numerically,  within the domain covering both an universal and semi-universal intervals, and with the use of the square root dependence,
inevitably results in a deviation from pure $\sqrt{\kappa\left(t_{*}-t\right)}$ law and an appearance of non-generic parameters $A$ and $c$. This observation is in good agreement with the experimental data \cite{key-8}. Remarkably, the boundary between these two intervals (which was found numerically) is of the order of the Schwarz criterion for reconnection ansatz \cite{key-11,key-12}.

The mutual friction affects also other features of reconnection processes. Thus,
under some conditions the vortex loops can shrink and die, before the collision of lines occurs. Additionally a lifetimes of significantly deformed vortices are close to the lifetime of the ideal single vortex ring.
We also observed the influence of friction on the pyramidal  structure of vortex line before reconnection. However, in the universal interval  this  pyramidal  structure is independent on the initial position of vortex
loops and temperature. The angles of pyramid don't change in the universal interval.

\end{document}